\documentclass[epj]{svjour}

\usepackage{amssymb}
\usepackage{amsmath}
\usepackage{graphicx}
\usepackage{hyphenat}

\let\textAA\AA
\def\AA{\text{\textAA}}

\begin{document}
\title{Relaxation dynamics in type-II superconductors with point-like and 
  correlated disorder}
\author{Ulrich Dobramysl \and Hiba Assi \and Michel Pleimling \and Uwe
  C. T\"{a}uber}
\institute{Department of Physics, Virginia Tech, Blacksburg, 
  Virginia 24061-0435\\
  \email{ulrich.dobramysl@vt.edu; tauber@vt.edu}}
\date{Received: date / Revised version: date}


\abstract{
  We employ an elastic line model to investigate the steady-state
  properties and non-equilibrium relaxation kinetics of magnetic
  vortex lines in disordered type-II superconductors using Langevin
  molecular dynamics (LMD). We extract the dependence of the mean
  vortex line velocity and gyration radius as well as the mean-square 
  displacement in the steady state on the driving current, and measure
  the vortex density and height autocorrelations in the aging regime.
  We study samples with either randomly distributed point-like or 
  columnar attractive pinning centers, which allows us to distinguish the 
  complex relaxation features of interacting flux lines subject to 
  extended vs. uncorrelated disorder. Additionally, we find that
  our new LMD findings match earlier Monte Carlo (MC) simulation data 
  well, verifying that these two microscopically quite distinct simulation
  methods lead to macroscopically very similar results for non-equilibrium 
  vortex matter.
}

\maketitle

\section{Introduction}
\label{sec:introduction}

Technical applications of type-II superconductors, especially in
high-field configurations, require effective pinning mechanisms to
prevent flux creep and flow and to thereby avoid dissipative losses. 
A disordered system of vortex lines at a finite temperature, subject to
pinning from randomly distributed point-like or correlated pinning sites, 
forms a remarkably complex system displaying a wealth of features. 
Naturally occurring weak point-like disorder already destroys the 
low-temperature Abrikosov lattice present in a clean system in favor of
a new Bragg glass phase with quasi long-range positional order. The 
first-order melting transition~\cite{Nelson1988,Nelson1989,Nelson1989a} 
is then replaced by a continuous transition into a vortex glass
phase~\cite{Fisher1989,Fisher1991,Feigelman1989,Nattermann1990}, in
which translational order is completely lost. Hence, vortex matter with 
weak, randomly placed, point-like disorder already shows a very rich 
phase diagram~\cite{Blatter1994,Banerjee2001}. The introduction
of correlated disorder, such as columnar pinning sites, results in a
strongly pinned Bose glass phase with localized vortex lines and a
diverging tilt modulus~\cite{Nelson1992,Lyuksyutov1992,Nelson1993},
which is accessible to analytical treatment via a mapping to the
propagation of bosons in imaginary time~\cite{Fisher1989a,Tauber1997}.

Since Struik's original investigation on physical aging in various
materials~\cite{Struik1978}, many glassy systems have been found to
show physical aging~\cite{Henkel2007}. More recent studies confirm 
that many other systems show the characteristics of glass-like 
relaxation and aging~\cite{Henkel2010,Cugliandolo2003,Henkel2008}. 
Experimentally, Du \textit{et al.} detected evidence of physical aging in 
disordered vortex matter by demonstrating that the voltage response 
of a 2H-NbSe$_2$ sample to a current pulse depends on the pulse
duration~\cite{Du2007}. Thin-film Monte Carlo relaxation and aging
studies of a coarse-grained two-dimensional model were performed by
Nicodemi and
Jensen~\cite{Nicodemi2002,Nicodemi2001a,Jensen2001,Nicodemi2001}.
Bustingorry, Cugliandolo and Dom\'{i}nguez investigated the relaxation
of vortex matter employing Langevin molecular dynamics (LMD) for a
three-dimensional line model, finding clear indications of physical aging 
in two-time quantities such as the density-density autocorrelation 
function, the linear susceptibility and the mean-square
displacement~\cite{Bustingorry2006,Bustingorry2007}.

In this paper, we report the results of a study comparing the
non-equilibrium relaxation kinetics of vortex lines in the presence of
randomly distributed point-like disorder and correlated columnar
disorder. The origin of point-like pinning sites can be either
naturally occurring, or artificially introduced crystal
defects. Similarly, correlated columnar disorder appears either as
line dislocations or in the form of material damage tracks stemming
from high-energy ion irradiation. It is well-established
experimentally that columnar disorder yields considerably enhanced
pinning efficiency over uncorrelated point-like
disorder~\cite{Civale1991}. Since linear pinning centers are extended
along one spatial dimension, one also expects profound differences in
the out-of-equilibrium relaxation of magnetic flux lines as compared
to samples with randomly distributed point pins.  Naturally, these
differences can only be addressed in a fully three-dimensional model
and numerical study.

We perform extensive Langevin molecular dynamics simulations on a
coarse-grained elastic line model of vortex matter. For non-equilibrium 
systems, it is important to compare and thus validate different 
microscopic realizations and simulation methods, in order to ascertain 
that the resulting macroscopic features stem from physical properties 
of the system and not from artifacts of the specific algorithm. We first 
present our LMD data on the steady-state vortex velocity and radius 
of gyration for driven flux lines subject to point-like or columnar 
disorder. To validate our simulation code, we compare these results for 
attractive point pins with earlier findings from Monte Carlo (MC) 
simulations. We proceed to systematically investigate the complex 
non-equilibrium relaxation behavior of a system of initially randomly 
distributed and perfectly straight vortex lines via various two-time 
observables. For point-like disorder, we again compare our novel LMD 
results with previously published MC data~\cite{Pleimling2011}. The 
main focus of this work is the distinct relaxation behavior of flux lines 
in the presence of randomly distributed columnar and point pinning 
centers.

Our paper is organized as follows: In the next Section we define and
explain the elastic line model as well as our LMD algorithm and discuss 
the values of the different system parameters. We then introduce the 
quantities that we are using in order to understand the non-equilibrium 
properties of interacting vortex lines in the presence of different 
types of attractive defects. Section~\ref{sec:steady-state-prop} is 
devoted to a discussion of the steady-state properties. We use this 
regime in order to validate the different algorithms used for the study 
of our system. Section~\ref{sec:relaxation} presents our numerical 
results. In a systematic study we disentangle the different effects due 
to the line tension, the vortex-vortex interaction, the pinning to defect 
sites, and the finiteness of the system. We discuss our main finding on 
how the different types of pinning centers, point-like and extended
columnar defects, affect the non-equilibrium relaxation process. Finally, 
we summarize our results in Section~\ref{sec:discussion}.

\section{Model and Simulation Protocol}
\label{sec:model-simul-prot}

\subsection{Effective Model Hamiltonian}
\label{sec:effect-model-hamilt}

We consider in the following a system of $N$ vortex lines in the
London limit, where the penetration depth is much larger than the
coherence length. In order to model the dynamics of the system we
employ a fully three-dimensional elastic line
description~\cite{Nelson1993,Das2003}.  The Hamiltonian of this system
is written as a functional of the vortex line trajectories
$\vec{r}_i(z)=\bigl(x_i(z),y_i(z)\bigr)$, where $z$ denotes the
direction of the applied external magnetic field, and consists of
three competing terms: the elastic line energy, the attractive
external potential due to disordered pinning sites, and the repulsive
vortex-vortex interactions:
\begin{equation}
  \label{eq:hamiltonian}
  \begin{split}
    H\left[\vec{r}_i(z)\right]=&\sum_{i=1}^N\int_0^Ldz\Biggl[
   \frac{\tilde{\epsilon}_1}{2}\left|\frac{d\vec{r}_i(z)}{dz}\right|^2
  + U_D(\vec{r}_i(z),z)\\
   &+ \frac{1}{2}\sum_{j\ne i}^NV(|\vec{r}_i(z)-\vec{r}_j(z)|) \Biggr] .
  \end{split}
\end{equation}

The elastic line stiffness or local tilt modulus is given by
$\tilde{\epsilon}_1 \approx
\Gamma^{-2}\epsilon_0\ln(\lambda_{ab}/\xi_{ab})$, where
$\Gamma^{-1}=M_{ab}/M_c$ represents the effective mass ratio or
anisotropy parameter, whereas $\lambda_{ab}$ and $\xi_{ab}$
respectively denote the London penetration depth and coherence length
in the $ab$ crystallographic plane. The in-plane vortex-vortex
interaction is given by $V(r)=2\epsilon_0K_0(r/\lambda_{ab})$, with
the zeroth-order modified Bessel function $K_0$ (essentially a
logarithmic repulsion that is exponentially screened at the scale
$\lambda$). In our simulations, the interaction is cut off at
$5\lambda_{ab}$ in order to avoid artifacts due to the periodic boundary
conditions. The $N_D$ pinning sites are modeled by randomly
distributed smooth potential wells of the form
\begin{equation*}
  \begin{split}
    U_D(\vec{r},z)\!=\!-\!\!\sum_{\alpha=1}^{N_D}\!\frac{b_0}{2}p\,\delta(z-z_\alpha)\!\!
   \left[1\!-\!\tanh\left(\!5\frac{\left|\vec{r}-\vec{r}_\alpha\right|-b_0}{b_0}
   \!\right)\right],
\end{split}
\end{equation*}
where $p \geq 0$ is the pinning potential strength, and $\vec{r}_\alpha$
and $z_\alpha$ indicate the in-plane and $z$ position of pinning site
$\alpha$. Lengths are measured in units of the pinning
potential width $b_0$. Energies are measured in units of
$\epsilon_0b_0$ with $\epsilon_0=(\phi_0/4\pi\lambda_{ab})^2$, and 
the magnetic flux quantum $\phi_0=hc/2e$.

\subsection{Langevin Molecular Dynamics}
\label{sec:lang-molec-dynam}

We employ a LMD algorithm to simulate the vortex line dynamics. To
this end, we discretize the system into layers along the $z$ axis. The
layer spacing corresponds to the crystal unit cell size $c_0$ along the
crystallographic c-direction~\cite{Das2003,Bullard2008}. Forces acting
on the vortex line vertices can then be derived from the properly
discretized version of the Hamiltonian (\ref{eq:hamiltonian}). We
proceed to numerically solve the (overdamped) Langevin equation
\begin{equation}
  \label{eq:langevin}
  \eta\frac{\partial\vec{r}_i(t,z)}{\partial t}=
  -\frac{\delta H[\vec{r}_i(t,z)]}{\delta\vec{r}_i(t,z)}+\vec{f}_i(t,z).
\end{equation}
with the Bardeen-Stephen viscous drag parameter $\eta$. 
The fast, microscopic degrees of freedom of the surrounding medium are 
\tolerance=1590
captured by thermal stochastic forcing, modeled as uncorrelated Gaussian 
white noise fulfilling $\left<\vec{f}_i(t,z)\right>=0$ and the Einstein relation
$\left<\vec{f}_i(t,z)\vec{f}_j(s,z')\right>=2\eta k_{\rm B}T\delta_{ij}
\delta(t-s)\delta(z-z')$, which guarantees that the system relaxes to 
thermal equilibrium with a canonical probability distribution 
$\propto e^{- H / k_{\rm B} T}$. The time integration is performed
via simple discretization of Eq.~\eqref{eq:langevin}~\cite{Brass1989}.

\subsection{Monte Carlo Algorithm}

In section~\ref{sec:steady-state-prop}, we will compare steady-state
results of systems of driven vortex lines generated by our LMD
algorithm to data stemming from MC simulations. The MC data was
obtained by applying the standard Metropolis update rule to the
Hamiltonian~\eqref{eq:hamiltonian}: A line element is picked at random
and made to jump in a random direction and (but truncated\footnote{A
  step size cut-off is necessary in order not to skip over pinning
  potential wells. See Refs.~\cite{Bullard2008,Klongcheongsan2010} for
  more information.}) distance. The ensuing change in the system's
energy $\Delta E$ is then evaluated and the Metropolis rule then
accepts the jump with a probability $P(\Delta E)=\min\{1,\exp(-\Delta
E/k_BT)\}$. One MC step is completed when exactly $NL$ line elements
have been selected.

\subsection{Material Parameters}
\label{sec:material-parameters}

We chose our simulation parameters to closely match the material
parameters of the ceramic high-$T_C$ type-II superconducting compound
YBa$_2$Cu$_3$O$_7$ (YBCO). The material is highly anisotropic with an
effective mass anisotropy ratio of $\Gamma^{-1}=1/5$. We set the
pinning center radius to $b_0=35\AA$ and measure simulation distances
in terms of this length. The in-plane London penetration depth and
coherence length are $\lambda_{ab}=34b_0\approx1200\AA$ and
$\xi_{ab}=0.3b_0\approx10.5\AA$ respectively. The vortex line energy
per unit length is $\epsilon_0\approx1.92\cdot
10^{-6}\operatorname{erg}/\operatorname{cm}$, hence the line tension
energy scale becomes $\tilde{\epsilon}_1\approx0.189\epsilon_0$. The
depth of the pinning center potential is set to $p=0.05\epsilon_0$,
except when noted differently. To fix the intrinsic simulation time scale, we 
set the Bardeen-Stephen viscous drag coefficient 
$\eta=\phi_0^2/2\pi\rho_nc^2\xi_{ab}^2 \approx10^{-10}
\operatorname{erg}\cdot\operatorname{s} / \operatorname{cm}^2$ to 
one (for the normal-state resistivity of YBCO near $T_C$, 
$\rho_n\approx500\mu\Omega\operatorname{cm}$, see table 1 in 
Ref.~\cite{Abdelhadi1994}), resulting in a basic time unit of
$t_0=\eta b_0/\epsilon_0\approx 18\operatorname{ps}$.

\subsection{Relaxation Simulation Protocol}
\label{sec:relax-simul-prot}

Throughout our study, the investigated systems contained $N=16$ vortex
lines with $L=640$ number of layers (except where noted differently). In 
the scenarios that include disorder, the number of pinning sites per layer is 
$N_D/L=1116$ which corresponds to a mean in-plane distance of $9b_0$ 
between pinning sites. In the case of randomly arranged point defects, 
the pinning site positions are chosen anew for each layer, whereas for 
columnar disorder, each layer repeats the pattern of the first layer's 
randomly chosen positions. The system size is set to 
$(16/\sqrt{3})\lambda_{ab}\times 8\lambda_{ab}$, with the aspect ratio 
chosen such that a clean system with interacting vortex lines reproducibly 
forms a hexagonal Abrikosov lattice configuration after equilibration. We 
employ periodic boundary conditions in the $x$ and $y$ directions, and 
free boundary conditions along the $z$ axis.

Our initial out-of-equilibrium condition consists of a system of
perfectly straight vortex lines, placed at random locations throughout
the computational domain. Since the vortex line elements do not yet
fluctuate (i.e. their distance from the vortex line mean in-plane
position is zero), the internal vortex line configuration effectively
is at zero temperature, hence, the start of the simulation at $t=0$ is
similar to an up-quench to a finite temperature
$T=10\operatorname{K}$. This is in contrast to their random spatial
distribution, which is equivalent to an infinite temperature. We then
let the system relax towards equilibrium until the waiting time $t=s$
(typically in the range of $2$ to $4096$) is reached, when we take a
snapshot of the system. We proceed to calculate various two-time
quantities (see section \ref{sec:measured-quantities} below) at
logarithmically-spaced time intervals with a simulation end time that
is ten times larger than the waiting time.

\subsection{Measured Quantities}
\label{sec:measured-quantities}

In the steady state of our driven flux line system, we measure the
mean vortex velocity $v$ by extracting the velocities of each
line element in the direction of the driving force $F_d$ from the time
stepping algorithm in LMD and average over all line elements in the
system. In MC, we take the average displacement in the direction of the
driving force $F_d$ over 30 MC steps and calculate the mean
velocity. Using Faraday's law, we can relate the vortex line mean
velocity to an induced electric field $\vec{E}=\vec{B}\times\vec{v}/c$,
which translates to a voltage drop across the sample. Similarly, the
driving force is related to an applied external current via the
Lorentz force $F_d=|\vec{j}\times\phi_0\vec{B}/B|$. Hence, a driving
force vs. mean vortex velocity graph is equivalent to experimentally
determined current-voltage (I-V) characteristics.

To quantify thermal spatial fluctuations along the vortex lines, we
compute the vortex line radius of gyration
$r_g=\sqrt{\left<(\vec{r}_{i,z}-\overline{\vec{r}}_i)^2\right>}$,
i.e. the root mean-square displacement from the lines' mean lateral
positions. The angular brackets again indicate an average over line
elements as well as noise and disorder realizations. This quantity is
expected to show a maximum at driving forces just below the depinning
transition.

To further accurately capture the relaxation and aging dynamics of
out-of-equilibrium disordered vortex line systems, we measure two-time
correlation quantities. Since we wish to compare the relaxation behavior
in LMD with previously measured MC data, we utilize the same two-time 
observables as in Ref.~\cite{Pleimling2011}: the 
\emph{height-height autocorrelation} function of the vortex lines, the 
two-time \emph{mean-square displacement} and the 
\emph{density-density autocorrelation} function. All these quantities 
depend on two times, labeled in the following as $s$ and $t$, with $s<t$.

The roughness or \emph{height-height autocorrelation} function of the 
vortex lines is defined by:
  \begin{equation}
    \label{eq:heightheightcorrelation}
    C(t,s)=\left<\left(\vec{r}_{i,z}(t)-\overline{\vec{r}}_i(t)\right)
   \left(\vec{r}_{i,z}(s)-\overline{\vec{r}}_i(s)\right)\right>,
  \end{equation}
where $\vec{r}_{i,z}(t)$ are the in-plane coordinates of line $i$ at layer 
$z$ at time $t$, $\overline{\vec{r}}_i(t)$ is the mean position of line $i$, 
and the averages are taken over all line elements as well as noise and 
disorder realizations. This quantity contains information about {\em local} 
thermal fluctuations of vortex line elements around the flux line's mean 
lateral position. In the case of free, non-interacting vortices, it can be 
mapped to the height correlation of growing one-dimensional interfaces 
(see Sec.~\ref{sec:free-nonint-lines}).

The two-time \emph{mean square displacement}, defined as
\begin{equation}
  \label{eq:meansqdisplacement}
  B(t,s)=\left<(\vec{r}_{i,z}(t)-\vec{r}_{i,z}(s))^2\right>\;,
\end{equation}
measures the average square distance between a vortex line element's
position at time $s$ and a subsequent time $t$. This quantity provides
data on the time evolution of the {\em global} structure of the vortex
line configuration, in addition to the same local information
contained in $C(t,s)$.

Finally, the two-time \emph{density-density autocorrelation} function
$C_v(t,s)$ is an observable that is measured by saving a snapshot of
the positions of all vortex line elements in the system at time $s$,
calculating the radial distance $r_i(t,s)$ each vortex line element
traveled between times $s$ and $t$ and determining the number $n_c$
of vortex line elements for which this distance is smaller than a
prescribed cutoff distance $r_i(t,s)<r_c$. The density autocorrelation
is then $C_v(t,s)=\left<n_c/NL\right>$. Throughout our study, the
cutoff distance is $r_c=0.05b_0$. The density autocorrelation also
contains information on the formation or decay of global structures;
thus we expect it to generally follow the behavior of $B(t,s)$.

\subsection{Comparison of Microscopic Algorithms}
\label{sec:comp-algor}

\tolerance=500
In order to simulate the dynamics of elastic lines in a disordered
medium, an appropriate microscopic algorithm has to be chosen. A
considerable amount of work has been done using Metropolis MC
implementations of the model described above and variations
thereof~\cite{Das2003,Bullard2008,Klongcheongsan2010}. 
Gotcheva \textit{et al.} investigated the differences between a 
Metropolis and a continuous-time MC algorithm for a system of flux lines 
on a discrete lattice and subject to varying temperature and driving
force~\cite{Gotcheva2004,Gotcheva2005}. The continuous-time update 
rule preserved positional order, while the Metropolis rule led to a 
disordered moving state, questioning the validity of the Metropolis  
algorithm for studies of driven vortex matter in lattice simulations. More 
recent studies demonstrated that positional order was preserved in 
off-lattice Metropolis MC of driven vortex matter~\cite{Bullard2008}.

It is crucial to investigate and compare different microscopic
implementations of algorithms such as Metropolis MC and LMD
simulations in a non-equilibrium setting. The choice of algorithm
might introduce spurious effects that cannot be predicted \textit{a
  priori}. In order to separate actual physical effects of the studied
elastic line model from these artifacts, we performed a careful
numerical comparison of LMD with earlier MC studies. Fast microscopic
degrees of freedom are modeled by the thermal force term in 
LMD, see Eq.~\eqref{eq:langevin}. In equilibrium the noise strength
is set by the Einstein relation (fluctuation-dissipation theorem, FDT). In 
out-of-equilibrium situations, there exists in general no FDT-equivalent 
that would uniquely determine the form and strength of the noise 
correlations. Since the large-scale and long-time characteristics of 
Langevin stochastic differential equations can be drastically influenced 
by the noise correlator properties, it is necessary to validate results by 
comparing to other numerical methods~\cite{Bullard2008}. In the first part 
of this article, we therefore compare results from a MC study for both 
steady-state and relaxation properties of the vortex line model to data 
generated by the LMD algorithm. It should be noted that a direct 
comparison of time scales is difficult since the length of a MC time step is 
a dynamically generated quantity, whereas in LMD the time step duration
is a function of the material parameters.

\section{Steady-state properties}
\label{sec:steady-state-prop}

\begin{figure}
  \centering
  \includegraphics[width=\columnwidth]{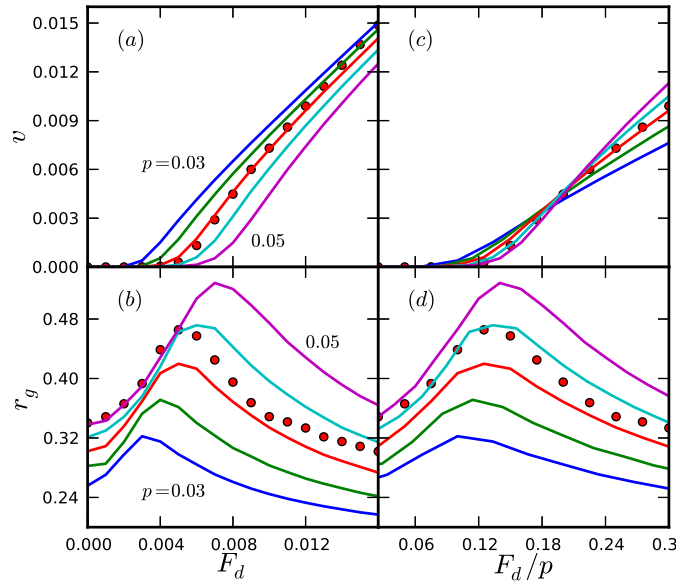}
  \caption{(Color online) Steady-state (a) velocity and (b) gyration
    radius of $N=16$ vortex lines with a length of $L=20$ elements as
    a function of the driving force $F_d$ in the presence of point pins 
    with varying disorder potential strength $p$. The red circles show 
    data from MC simulations with $p=0.05\epsilon_0$, while the other 
    graphs display LMD simulation data with $p$ ranging from 
    $0.03\epsilon_0$ to $0.05\epsilon_0$ in steps of $0.005\epsilon_0$. 
    In (c) and (d), the force axis is rescaled with the disorder potential
    strength. The velocity curves in (c) cross at $F_d/p\approx0.19$
    while the gyration radius maxima in (d) align around
    $F_d/p\approx0.12$, with a slight bias towards higher $F_d/p$ for
    higher values of $p$.}
  \label{fig:steady-state-ivgyrrad-pointlike}
\end{figure}
We first employ LMD simulations for interacting flux lines in the
presence of point-like disorder with disorder potential strength $p$,
subject to a driving force $F_d$ stemming, via the Lorentz force, from
an external current. Results for the steady-state velocity and
gyration radius are gathered in
Fig. ~\ref{fig:steady-state-ivgyrrad-pointlike}. The red dots in
Fig.~\ref{fig:steady-state-ivgyrrad-pointlike} indicate MC-generated
data, while the solid lines were produced using LMD with different
values of $p$. It is quite clear from
Fig.~\ref{fig:steady-state-ivgyrrad-pointlike}(a) that a pinning
potential strength $p=0.05\epsilon_0$ in MC corresponds to
$p\approx0.04\epsilon_0$ in LMD.  In MC, vortex line elements test a
region with a radius of $0.25b_0$ around their current position for
possible jump targets. It is conceivable that the pin energy barrier
appears a bit smoother in MC since its width falls into the same
length scale, which leads to the observed renormalization of the
pinning potential strength. The slightly higher maximum of the MC
gyration radius data for $p=0.05\epsilon_0$ over the corresponding LMD
curve with $p=0.04\epsilon_0$ in
Fig.~\ref{fig:steady-state-ivgyrrad-pointlike}(b) supports this
argument, since vortex line elements are most likely trapped at a
pinning site until they escape via a single jump. It is much less
probable for any line element to escape the pin via multiple
successive jumps. In LMD on the other hand, the thermal force is only
an added component on top of the (in this case stronger) driving and
elastic tension forces. Hence, LMD yields smooth escape trajectories
out of a point pin's binding potential, and the vortex line is
consequently less rough, resulting in a smaller radius of gyration.

Figures~\ref{fig:steady-state-ivgyrrad-pointlike}(c) and (d) show that
the depinning force scales roughly linear with the pinning potential
strength $p$, as expected~\cite{Nelson1993}. The vortex velocity
curves cross at $F_d/p\approx0.19$, while the gyration radii have
their maxima around $F_d/p\approx0.12$, with a slight systematic shift
to lower $F_d/p$ for smaller pinning strength values $p$. It should be
noted that a true continuous non-equilibrium depinning phase
transition occurs only at $T=0$ (and in the thermodynamic limit). The
scaling behavior of the velocity of driven vortex lines near the
critical depinning force in the presence of point-like disorder has
been explored by Luo and Hu~\cite{Luo2007}. The gyration radius
maximum in the vicinity of the critical depinning force may
consequently be understood as the thermally rounded remnant of this
zero-temperature phase transition~\cite{Fisher1998}.

\begin{figure}
  \centering
  \includegraphics[width=\columnwidth]{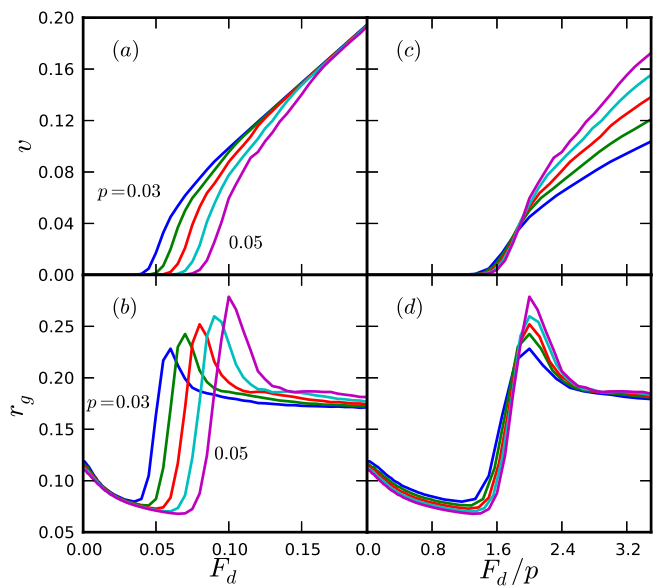}
  \caption{(Color online) Steady-state (a) velocity and (b) gyration
    radius of $N=16$ vortex lines with $L=20$ as a function of the
    driving force $F_d$ in the presence of columnar disorder, with a
    disorder potential strength varying between $p=0.03\epsilon_0$ and
    $p=0.05\epsilon_0$ in steps of $0.005\epsilon_0$. As expected, the
    radius of gyration displays quite different behavior for columnar
    pins as compared to point pinning sites, compare
    Fig.~\ref{fig:steady-state-ivgyrrad-pointlike}(b).  In (c) and (d), the 
    force axis is rescaled with the disorder potential strength. The velocity 
    curves in (c) cross at $F_d/p\approx1.9$, while the gyration radius 
    maxima in (d) align around $F_d/p\approx2$.}
  \label{fig:steady-state-ivgyrrad-columnar}
\end{figure}
Figures~\ref{fig:steady-state-ivgyrrad-columnar}(a) and (b) display
the mean velocity $v$ and the radius of gyration $r_g$ as a function
of the driving force $F_d$ of vortices subject to randomly distributed
columnar pinning sites. Correlated disorder is much more effective at
pinning flux lines than point disorder~\cite{Civale1991}. This is
reflected in a critical depinning force that is about an order of
magnitude higher for columnar defects than for uncorrelated point pins
of the same strength $p$ (per layer; results shown in
Fig.~\ref{fig:steady-state-ivgyrrad-pointlike}). In
Figs.~\ref{fig:steady-state-ivgyrrad-columnar}(c) and (d) the driving
force is again scaled with the pin strength $p$. The vortex velocity
curves cross at $F_d/p\approx1.9$, which is indeed a factor of $10$
larger compared to point pins. In the presence of columnar defects, a 
single flux line may be in one of the following four 
configurations:~\cite{Nelson1993} (i) unpinned, located away from any
pinning sites; (ii) trapped at a single columnar pin for its entire
length; (iii) forming a vortex half-loop where the elastic line is
trapped at a single columnar defect, with the exception of an unpinned
section that extends away from the defect line: Depending on the
relative strengths of the driving force, thermal noise, and the
pinning potential, the unpinned part may either expand or
retract. This state represents a short-lived saddle-point configuration;
(iv) forming single or double kinks by being simultaneously trapped at
two adjacent pinning columns. This state is rather long-lived but will
ultimately decay into either the unpinned or completely trapped state.

The gyration radius indicates which configurations are typically
assumed by the vortex lines. For $F_d=0$, most vortex lines are fully
trapped. The radius of gyration of a trapped line is restricted by the 
pinning potential extension $b_0$, hence we observe a marked
reduction in the value of $r_g$ with columnar pins over free, unbound
lines. With increasing but below-critical $F_d$, $r_g$ decreases since
the mean position of trapped vortex lines shifts from the center of
the pinning site, which further constrains fluctuations. In the vicinity of
the critical depinning force, in the flux creep regime, $r_g$ rises 
sharply due to the formation of half-loops, single-, and double-kinks. 
Near the transition to free-flowing flux lines, $r_g$ develops a 
maximum and gradually decreases for even higher $F_d$. In this 
state, vortex motion is slightly restricted by pinning centers, but the
flux lines move essentially unimpeded, and the radius of gyration 
approaches its unbound value.

\section{Relaxation processes}
\label{sec:relaxation}

In order to study the relaxation dynamics and possible aging scaling of 
a system of vortex lines in various scenarios, we follow the procedures
outlined in Ref.~\cite{Pleimling2011}. As a test case for our
simulation code, we first investigate a system of non-interacting
lines without disorder, which can be mapped to the one-dimensional
Edwards-Wilkinson (EW) interface growth model. We then proceed to
non\hyp{}interacting lines in the presence of point-like disorder with two
different pin potential strengths, where the short-time behavior is
similar to the clean system while the long-time relaxation is modified
by the attractive defects. To disentangle the effects of the mutual
vortex repulsion from the disorder influence, we next study a system 
of interacting vortex lines without pinning sites. Subsequently, we 
present data on the full system of interacting vortex lines in the 
presence of point pinning centers. We then point out the differences
in the relaxation kinetics in systems with point and correlated extended 
pins by investigating both non-interacting and interacting flux lines in 
the presence of columnar defects. Finally, we discuss finite-size
effects due to short vortex lengths for both types of pinning sites.

In each of the scenarios presented below, we first look at the
relaxation of the single-time mean-square displacement $B(t,0)$, the 
time-dependent squared radius of gyration, $r_g^2(t)$, and the
associated effective exponents $\beta_B(t)=d\ln B(t,0)/d\ln t$ and
$\beta_h(t)=d\ln r_g^2(t)/d\ln t$. The mean-square displacement 
predominantly probes changes in the average positions of single 
vortices. Hence we use the average of $\beta_B$ over an appropriate 
time interval as the aging exponent that we utilize to achieve 
(approximate) data collapse of the two-time global mean-square 
displacement $B(t,s)$ and density autocorrelation function $C_v(t,s)$. 
The time-dependent radius of gyration describes internal thermal 
vortex line fluctuations and enables us to compute an averaged 
scaling exponent $\beta_h$ that can be used for obtaining data 
collapse of the two-time height autocorrelation function $C(t,s)$.

\subsection{Free Non-interacting Vortex Lines}
\label{sec:free-nonint-lines}

The thermal fluctuations of the segment locations of free 
(non-interacting and not subject to disorder) directed elastic lines 
around their mean in-plane position can be map\-ped to the problem of a
one-dimensional interface growing via random deposition. The
continuous version of this growth model is described by the stochastic
Edwards-Wilkinson (EW) equation~\cite{Edwards1982}. The temporal
evolution of the interface height relative to its mean height is governed
by a diffusive term as well as a random noise term. Hence, it may be
described mesoscopically by a linear Langevin equation,
\begin{equation}
  \label{eq:edwards-wilkinson}
  \frac{\partial h(z,t)}{\partial t}=
  \nu\frac{\partial^2 h(z,t)}{\partial z^2}+\eta(z,t)\;,
\end{equation}
where $\nu$ is the diffusive strength and $\eta(z,t)$ represents 
thermal white noise with zero mean and second moment
$\left<\eta(z,t)\eta(z',s)\right>=2k_{\rm B}T\nu\delta(z-z')\delta(t-s)$ that
satisfies Einstein's relation. The temperature $T$ enters through the 
noise strength.

The linear nature of Eq.~\eqref{eq:edwards-wilkinson} makes it
possible to arrive at analytical expressions for various two-time 
quantities~\cite{Rothlein2007,Bustingorry2007,Bustingorry2007a,Chou2010}. In
particular, the solution for the two-time height-height autocorrelation 
function in the correlated growth regime reads~\cite{Rothlein2007},
\begin{equation}
  \label{eq:1}
  C(t,s)=C_0s^{1/2}\left(\left[\frac{t}{s}+1\right]^{1/2}
  -\left[\frac{t}{s}-1\right]^{1/2}\right)\;.
\end{equation}
\tolerance=2000
Comparing with the general scaling form $C(t,s)=s^{-b}f_C(t/s)$ (with
scaling function $f_C$), this predicts the universal aging exponent
$b=1/2$ in the EW regime~\cite{Pleimling2011,Rothlein2007}. The mean
square displacement follows a similar scaling form
$B(t,s)=s^{-b}f_B(t/s)$, while the density autocorrelation empirically
scales as $C_v(t,s)=s^{b}f_{C_v}(t/s)$.

\begin{figure}
  \centering
  \includegraphics[width=\columnwidth]{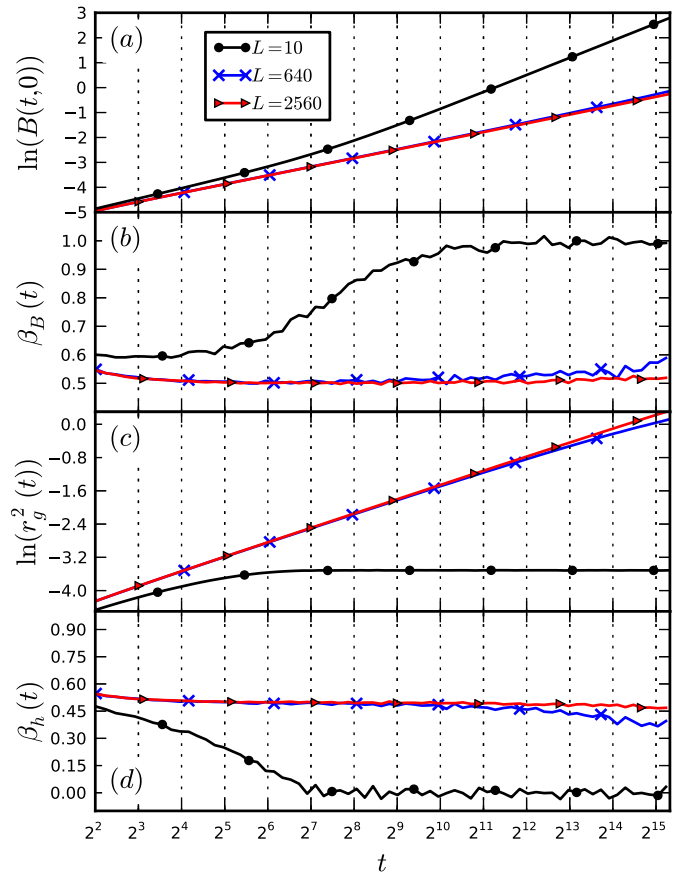}
  \caption{(Color online) Relaxation behavior of (a) the vortex line
    mean-square displacement $B(t,0)$, (c) the squared gyration radius
    $r_g^2(t)$, and (b,d) the associated effective exponents $\beta_B$
    and $\beta_h$ over time for free non-interacting flux lines with
    length $L=10$ ($\bullet$ marker), $L=640$ ($\times$
    marker) and $L=2560$ ($\blacktriangleright$ marker) averaged over
    at least 1000 realizations.}
  \label{fig:relaxation-free-line}
\end{figure}
\begin{figure}
  \centering
  \includegraphics[width=\columnwidth]{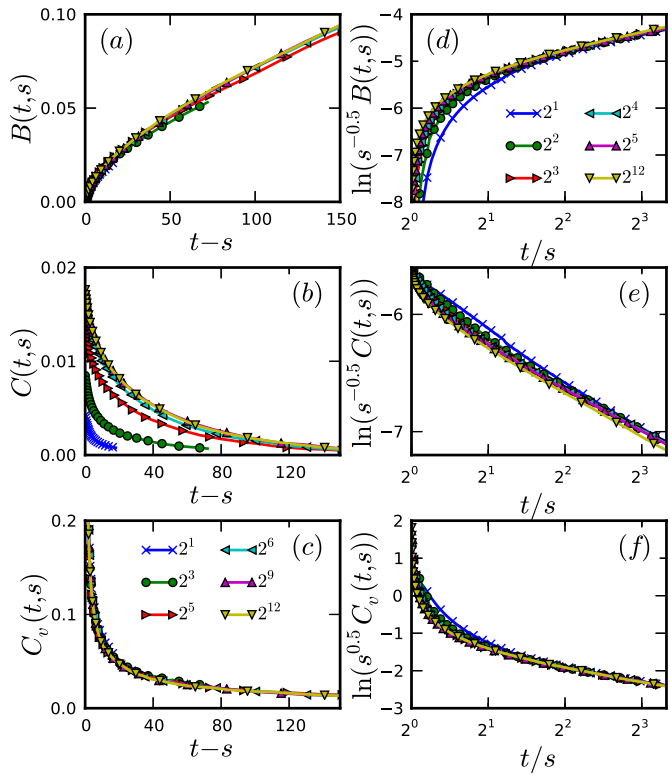}
  \caption{(Color online) Relaxation of two-time quantities in a
    system of non-interacting flux lines without disorder for a line
    length of (a-c) $L=10$ and (d-f) $L=2560$, averaged over $800$
    different noise realizations. The sub-figures show (a,d) the
    mean-square displacement $B(t,s)$, (b,e) the height autocorrelation 
    function $C(t,s)$, and (c,f) the density autocorrelation function 
    $C_v(t,s)$ as a function of (a-c) $t-s$ and (d-f) $t/s$. Waiting times 
    range from $s=2$ to $2^{12}=4096$. For $L=10$, the system 
    rapidly reaches equilibrium. The height autocorrelation function in (b)
    displays a short time window ($s<64$) where it explicitly depends on
    $s$ and time translation invariance is broken. For $L=2560$ and 
    $s>16$, all quantities show aging and dynamical scaling with the EW 
    exponent $b=0.5$.}
  \label{fig:p_free_line}
\end{figure}
The relaxation of the observables $B(t,0)$, $r_g^2(t)$ and the three 
two-time correlation functions $B(t,s)$, $C(t,s)$, and $C_v(t,s)$ in this 
scenario as observed in our LMD simulations is presented in 
Figs.~\ref{fig:relaxation-free-line} and~\ref{fig:p_free_line},
respectively. For a very thin system with $L=10$, we immediately see 
from the time evolution of the effective exponent $\beta_B(t)$ [solid 
line with black circles in Fig.~\ref{fig:relaxation-free-line}(b)] that the 
system starts to cross over into equilibrium where $B(t,0)\sim t$, 
$\beta_B\to1$ for early times $t>2^5$, and it is truly equilibrated after 
$t>2^{10}$. This is also visible in the unscaled two-time quantities 
$B(t,s)$ and $C_v(t,s)$ displayed in Figs.~\ref{fig:p_free_line}(a,c): 
For $s>2^4$, the data of $B(t,s)$ for different waiting times $s$ fall
onto a single master curve, which indicates the recovery of time 
translation invariance. In contrast, the data for $C_v(t,s)$ collapse for 
all waiting times $s$. The difference in the onset of data collapse in 
these two quantities is caused by local thermal fluctuations contributing 
to the mean-square displacement, whereas short-scale variations are 
effectively averaged out due to the finite cutoff radius in the density
autocorrelation\footnote{See the description of the algorithm for 
  calculating $C_v(t,s)$ in Sec.~\ref{sec:measured-quantities} and 
  Ref.~\cite{Pleimling2011} for more information.}. The effective 
gyration radius exponent $\beta_h(t)$, displayed in 
Fig.~\ref{fig:relaxation-free-line}(d), decreases from the start and 
eventually reaches zero around $t\approx2^7$. For such a short flux 
line length, the crossover from the EW regime to the saturated
(equilibrated) regime happens at very early times\footnote{See
  Ref.~\cite{Chou2009} for a discussion of different EW regimes 
  and the length dependence of the crossovers.}. The two-time
height-height autocorrelation function $C(t,s)$, plotted in
Fig.~\ref{fig:p_free_line}(b), shows data collapse for waiting times 
$s>2^7$, which reflects the time evolution of $\beta_h(t)$.

The extended, bulk-like system with $L=2560$ exhibits much slower 
relaxation and hence enables us to study the dynamical aging scaling 
regime. The effective exponents $\beta_B(t)$ and $\beta_h(t)$ [solid 
lines with black triangles in Figs.~\ref{fig:relaxation-free-line}(b,d)] 
show the remnants of a crossover for short times, while staying at a 
value of approximately $0.5$ for $t>2^4$. 
Figures~\ref{fig:p_free_line}(d-f) depict $B(t,s)$, $C(t,s)$ and
$C_v(t,s)$ respectively as functions of $t/s$ and scaled with
appropriately chosen exponents of the waiting time. The data for all
three quantities show data collapse for $s>2^4$ with exponent 
$b=0.5$, indicating dynamical scaling and hence full aging in 
this time regime. The aging exponent of the two-time height-height 
autocorrelation function coincides with the predicted EW value from 
Eq.~\eqref{eq:1}.

Except for the early-time crossover between the EW and saturation
regimes, which is not visible in Fig.~3 in Ref.~\cite{Pleimling2011}, 
our findings produced via LMD simulations are in complete agreement 
with the MC data for both thin and extended ``bulk'' systems.

\subsection{Non-interacting Vortices with Point Disorder}
\label{sec:nonint-lines-disorder}

\begin{figure}
  \centering
  \includegraphics[width=\columnwidth]{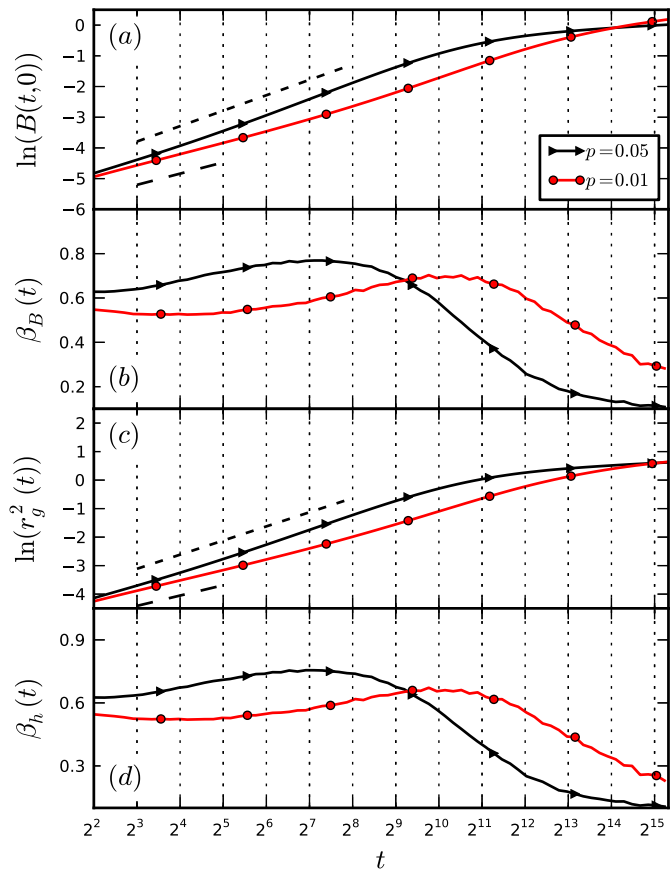}
  \caption{(Color online) Relaxation behavior of (a) the flux line 
    mean-square displacement $B(t,0)$, (c) the squared gyration 
    radius $r_g^2(t)$, and (b, d) the associated effective exponents 
    $\beta_B$ and $\beta_h$ over time for non-interacting vortices 
    subject to point pins with potential strength $p=0.01\epsilon_0$ 
    ($\bullet$ marker) and $p=0.05\epsilon_0$ ($\blacktriangleright$), 
    averaged over $1000$ realizations. For $p=0.01\epsilon_0$, the 
    dashed lines (below the curves) indicate the power laws with mean 
    effective exponents $\overline{\beta_B}\approx0.527\pm0.003$ 
    in (a) and $\overline{\beta_h}\approx0.523\pm0.002$ in (c) over 
    the range $2^3\le t\le 2^{5}$. Similarly for $p=0.05\epsilon_0$, 
    the dashed lines (above the curves) show the power laws with mean 
    effective exponents $\overline{\beta_B}\approx0.725\pm0.041$ 
    and $\overline{\beta_h}\approx0.716\pm0.037$ over the range  
    $2^3\le t\le2^8$.}
  \label{fig:relaxation-noninter}
\end{figure}
\begin{figure}
  \centering
  \includegraphics[width=\columnwidth]{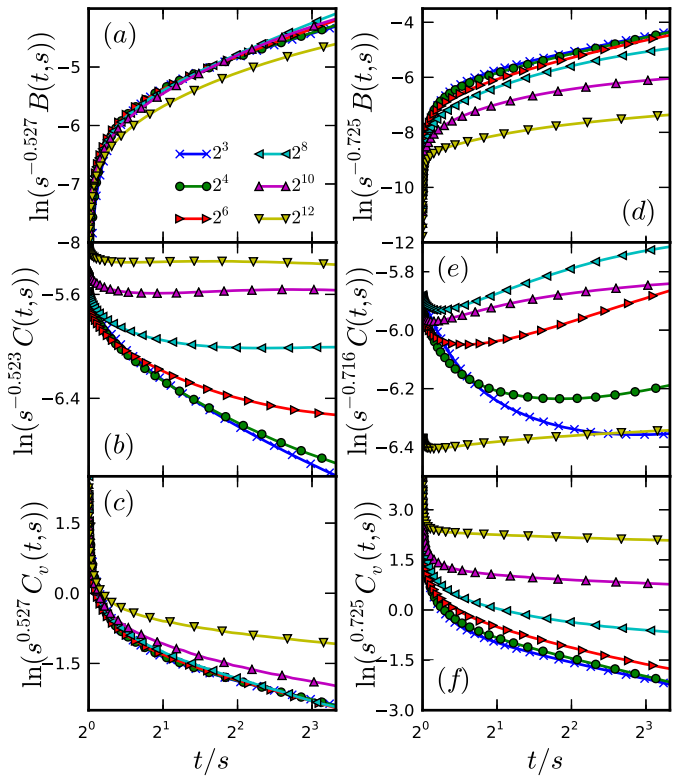}
  \caption{(Color online) Relaxation of (a,d) the mean-square
    displacement, (b,e) the height autocorrelation function, and (c,f) 
    the density autocorrelation in a system of non-interacting vortex 
    lines of length $L=640$, subject to randomly distributed point pins 
    with a potential depth of (a-c) $p=0.01\epsilon_0$ and (d-f) 
    $p=0.05\epsilon_0$; data averaged over $1000$ realizations. Time 
    translation invariance is broken throughout the simulation time 
    window. For both pinning strengths, dynamical scaling for the 
    mean-square displacement and the density autocorrelation 
    approximately holds in an intermediate range of waiting times $s$, 
    with the mean effective exponents inferred from  
    Fig.~\ref{fig:relaxation-noninter}.}
  \label{fig:p_noninter}
\end{figure}
We now proceed to add point-like disorder with pinning potential
strengths $p=0.01\epsilon_0$ and $p=0.05\epsilon_0$ to a system 
of non-interacting vortex lines, see Fig.~\ref{fig:relaxation-noninter}. 
The equilibrium configuration at low temperatures constitutes an
extremely dilute vortex glass. For the smaller defect strength of 
$p=0.01\epsilon_0$ there exists an intermediate time regime 
$2^3<t<2^7$ during which both effective exponents $\beta_B(t)$ 
and $\beta_h(t)$ are fairly constant before developing a maximum 
around $t=2^{10}$ with a subsequent crossover into a frozen state 
with $\beta_B,\beta_h\to 0$ where the vortex lines are firmly bound 
to the point defects [solid lines with black circles in 
Fig.~\ref{fig:relaxation-noninter}(b,c)]. The slight downward slope at 
early times is a remnant of the crossover from the random-noise into 
the EW regime of free flux lines. In fact, the exponent values at times 
$t<2^7$ approximately equal those of the disorder-free system, 
indicating that the time evolution of the mean lateral vortex position is 
essentially the same as for free lines prior to the disorder effects 
becoming noticeable. At later times, vortex movement starts to 
become affected by the attractive pinning sites, which temporarily 
accelerates the relaxation kinetics as the vortices are drawn into 
potential wells, before flux line motion becomes at last frozen at the 
defects.

Figures~\ref{fig:p_noninter}(a-c) show the resulting relaxation of the
two-time mean square displacement $B(t,s)$, the height-height
autocorrelation function $C(t,s)$ and the density-density
autocorrelation function $C_v(t,s)$ for the system with pinning
strength $p=0.01\epsilon_0$. The global quantities $B$ and $C_v$ 
are scaled using the aging exponent $b=0.527$ taken from the 
average effective exponent $\beta_B$ over the short- to 
intermediate-time region $2^3<t<2^5$, where $\beta_B$ is roughly 
constant. These two-time autocorrelation functions yield approximate 
dynamical scaling for waiting times in this time regime, which further 
supports the interpretation that pinning sites are essentially irrelevant 
for the motion of the vortex line mean lateral positions during the early 
stages of the relaxation process.

In the short-time regime $t<2^5$, the squared gyration radius
$r_g^2(t)$ in Fig.~\ref{fig:relaxation-noninter}(c,d) is also described
by a power law with an approximate effective exponent 
$\overline{\beta_h}\approx0.523$. Using this value as the aging
exponent for the scaled two-time height-height autocorrelation 
function $C(t,s)$ in Fig.~\ref{fig:p_noninter}(b) reveals approximate
dynamical scaling in this time window. However, stronger deviations
from the free-line behavior are observed for $C(t,s)$ than for the 
other quantities. For larger times $t>2^6$ correlations become 
increasingly longer-lived, since vortex line elements become trapped at 
pinning sites. Hence, the influence of weak point defects is observed
mainly in the fluctuations of flux line elements, whereas the movements 
of their mean lateral positions are hardly modified\footnote{The
  supplementary movie ``NonInteractWeakLargePinsNearby.mp4'' 
  shows an example realization for this behavior.}.

For a larger pin strength $p=0.05\epsilon_0$, the effective exponent
maxima in Figs.~\ref{fig:relaxation-noninter}(b,d) develop earlier,
near $t=2^{7}$, and there appears no region with approximately
constant exponents. For $t>2^{11}$, the system crosses over into a
regime where the vortex line configuration appears to become frozen at
the pins. This is reflected also in the two-time quantities, where we
only see approximate data collapse for $B$ and $C_v(t,s)$ for waiting
times $s<2^6$ [Figs.~\ref{fig:p_noninter}(d,f)]. The two-time height
autocorrelation function $C(t,s)$ in Fig.~\ref{fig:p_noninter}(e)
yields interesting non-monotonic behavior for $s>2^3$, where
correlations actually increase again after developing a minimum. We
interpret this effect as a rather complicated cross-over effect caused
by competition of repulsive, pinning, and elastic forces. Initially,
the vortex lines locate nearby pinning sites and are drawn into their
attractive potential wells. This yields accelerated super-diffusive
motion as indicated by the exponent maximum in
Fig.~\ref{fig:relaxation-noninter}(d), until the elastic interaction
restricts further exploration of the configuration space and leads to
the subsequent decrease of the effective exponent. This behavior is
not apparent in the case of $p=0.01$ owing to our choice of waiting
times. The exponent maximum in this situation would occur at a much
later time and hence the highest waiting time $s=2^{12}$ does not yet
display any non-monotonic features.

Elastic manifolds subject to disorder can be characterized by means of
the roughness exponent $\chi$ which is defined via the height-height 
correlation function along the manifold dimensions~\cite{Nattermann2000} 
(here, the contour length of the directed lines)
\begin{equation}
  \label{eq:2}
  C(z-z')=\left<[\vec{r}(z)-\vec{r}(z')]^2\right>\sim|z-z'|^{2\chi}\,.
\end{equation}
In the case of a dilute flux line system, for which mutual interactions 
may be neglected, and free of disorder, thermal fluctuations lead to the
EW roughness exponent $\chi=0.5$, in agreement with our numerical
observations. In the dilute vortex glass phase with point-like disorder, 
renormalization group analysis of manifolds subject to Gaussian
disorder~\cite{Laessig1998,Nattermann2000} predicts a roughness 
exponent $\chi=5/8$. Our LMD simulations yield a distance-dependent 
effective roughness exponent in the range of $0.5<\chi_{eff}<0.8$ for 
a system of non-interacting vortex lines with point-like pinning sites. It 
should be emphasized, though, that in our model the pinning sites are 
exclusively attractive. It turns out that the out-of-equilibrium relaxation 
behavior is considerably different for directed lines subject to a mixture 
of attractive and repulsive pins: One then actually observes simple aging, 
albeit with non-universal scaling exponents that depend on temperature 
as well as pinning strength~\cite{Iguain2009,Pleimling2011}. 

\subsection{Interacting Vortex Lines without Disorder}
\label{sec:free-inter-lines}

\begin{figure}
  \centering
  \includegraphics[width=\columnwidth]{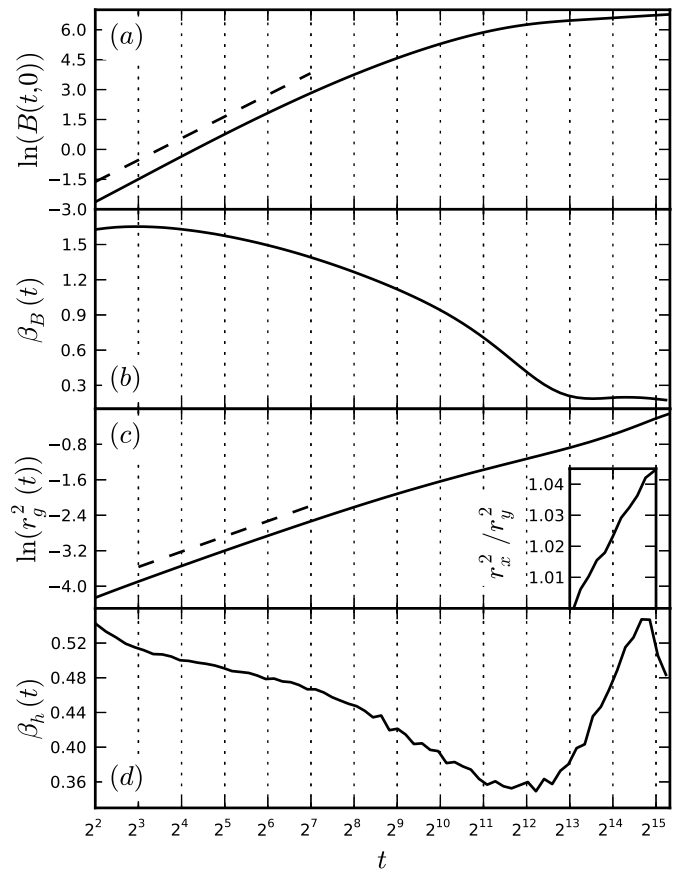}
  \caption{Relaxation behavior of (a) the flux line mean-square
    displacement $B(t,0)$, (c) the squared gyration radius $r_g^2(t)$,
    and (b, d) the associated effective exponents $\beta_B$ and 
    $\beta_h$ over time for interacting vortices in a system without 
    pinning centers, averaged over $5000$ realizations. The dashed lines 
    indicate power laws with the mean effective exponents (b) 
    $\overline{\beta_B}\approx1.57\pm0.08$ (averaged over the time 
    range $2^2\le t\le 2^6$) and (d) 
    $\overline{\beta_h}\approx0.50\pm0.02$ (averaged over 
    $2^3\le t\le 2^{7}$). The inset in (c) shows the ratio of the $x$ and
    $y$ components of the radius of gyration for $t>2^{12}$.}
  \label{fig:relaxation-interclean}
\end{figure}
\begin{figure}
  \centering
  \includegraphics[width=\columnwidth]{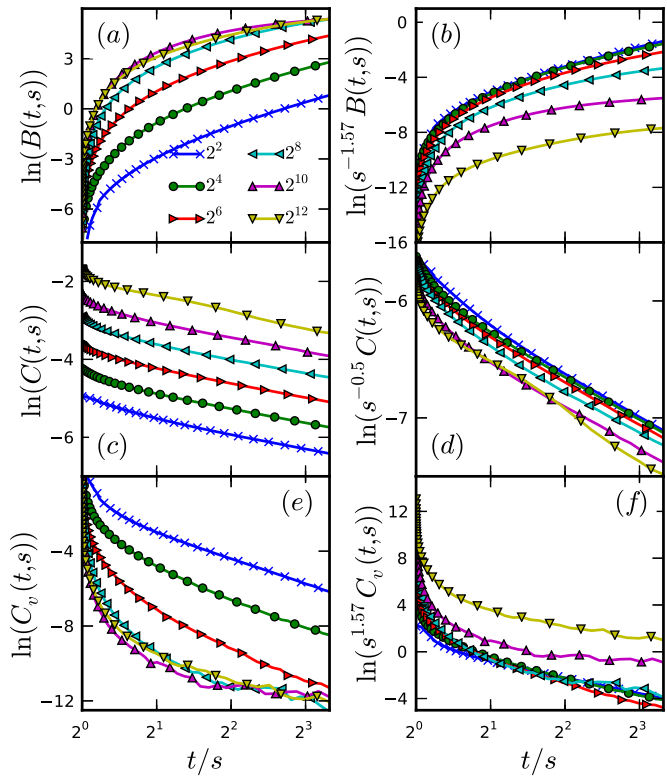}
  \caption{(Color online) Relaxation of (a,d) the mean-square
    displacement, (b,e) the height autocorrelation function, and (c,f) the 
    density autocorrelation in a system of interacting flux lines without 
    disorder and $L=640$; data averaged over $800$ realizations. The 
    left-hand panels show the unscaled log-log data, whereas data on 
    the right-hand side are scaled by the waiting time $s$ using the 
    mean exponents from Fig.~\ref{fig:relaxation-interclean}.}
  \label{fig:p_interclean}
\end{figure}
To disentangle the effects of mutual flux line repulsion from the
influence of disordered point pinning sites, we next study the
non-equilibrium relaxation of a clean system of interacting vortex
lines.  Fig.~\ref{fig:relaxation-interclean} shows the relaxation of
the mean-square displacement, the radius of gyration, and the
associated effective exponents in this scenario, whereas
Fig.~\ref{fig:p_interclean} displays the behavior of the three
two-time autocorrelations. One may immediately identify striking
differences between non-interacting and mutually repelling vortex
lines in the relaxation of $B(t,s)$ and its associated effective
exponent $\beta_B(t)$; compare
Figs.~\ref{fig:relaxation-free-line}(a,b) and
Figs.~\ref{fig:relaxation-interclean}(a,b). The initially large value
of $\beta_B$ in Fig.~\ref{fig:relaxation-interclean}(b) can be traced
to the rapid formation of long-range order due to repulsive vortex
interactions. This effect stems from our choice of initial conditions,
where the vortex lines are randomly distributed throughout the
system. Owing to the initially non-ideal spacing, mutual repulsion 
leads to fast vortex motion and thus to a large value of $\beta_B$. As 
soon as an optimal arrangement (in this case the Abrikosov lattice) is
reached, flux lines perform confined random walks due to the efficient
caging from neighboring vortices, which eventually leads to a low
effective exponent $\beta_B\approx0.2$ for $t>2^{12}$. The data
collapse in Fig.~\ref{fig:p_interclean}(b) shows that an averaged
$\overline{\beta_B}\approx1.6$ may serve as the effective aging
exponent for the two-time mean square displacement $B(t,s)$ for short
waiting times $s<2^6$.

The difference between interacting and non-interacting systems does not
appear as drastic for the squared radius of gyration $r_g^2(t)$ as for
$B(t,0)$, compare Figs.~\ref{fig:relaxation-free-line}(c,d) and
Figs.~\ref{fig:relaxation-interclean}(c,d). After $t>2^{10}$, the
Abrikosov lattice starts to form, and the repulsive forces due to
neighboring vortex lines increasingly suppress transverse flux line 
wandering. For $t>2^{12}$, the gyration radius components  along the 
$x$ and $y$ directions assume slightly different values owing to the 
anisotropic hexagonal vortex line arrangement, which in our rectangular 
system is always oriented along the $x$ direction; see the inset in 
Fig.~\ref{fig:relaxation-interclean}(c). For small waiting times $s<2^6$,
the height autocorrelation data can be collapsed with the EW aging
scaling exponent $b=0.5$.

\subsection{Interacting Vortex Lines with Point Disorder}
\label{sec:inter-lines-disorder}

\begin{figure}
  \centering
  \includegraphics[width=\columnwidth]{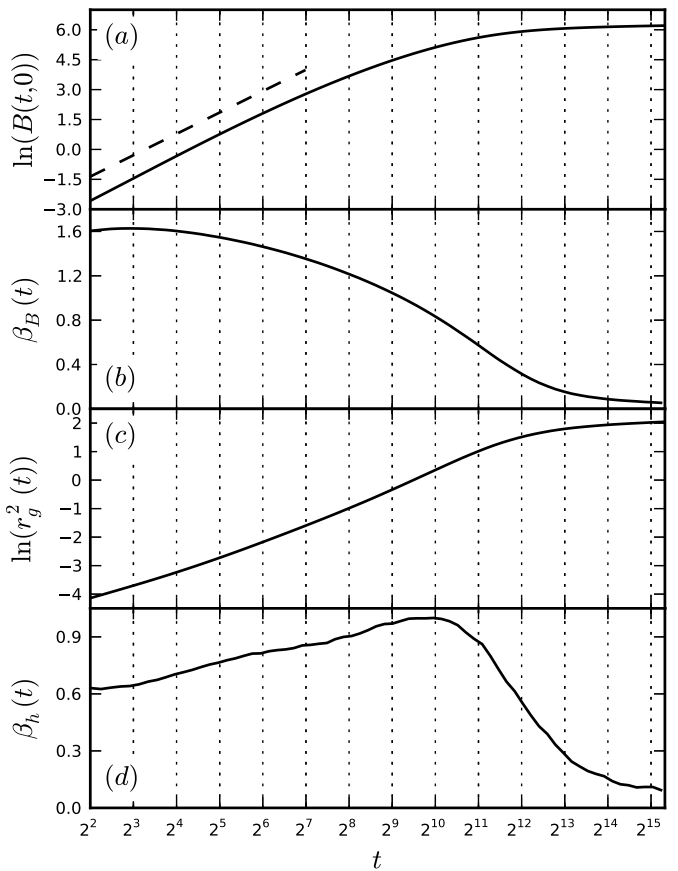}
  \caption{Relaxation behavior of (a) the flux line mean-square
    displacement $B(t,0)$, (c) the squared gyration radius $r_g^2(t)$,
    and (b, d) the associated effective exponents $\beta_B$ and $\beta_h$ 
    over time for interacting vortices in a system with point-like disorder of 
    strength $p=0.05\epsilon_0$, averaged over $1000$ realizations. The 
    dashed line in (a) indicates a power law with the mean effective 
    exponent $\overline{\beta_B}\approx1.54\pm0.08$ (averaged over the 
    time interval $2^2\le t\le 2^7$).}
  \label{fig:relaxation-fullsystem}
\end{figure}
\begin{figure}
  \centering
  \includegraphics[width=\columnwidth]{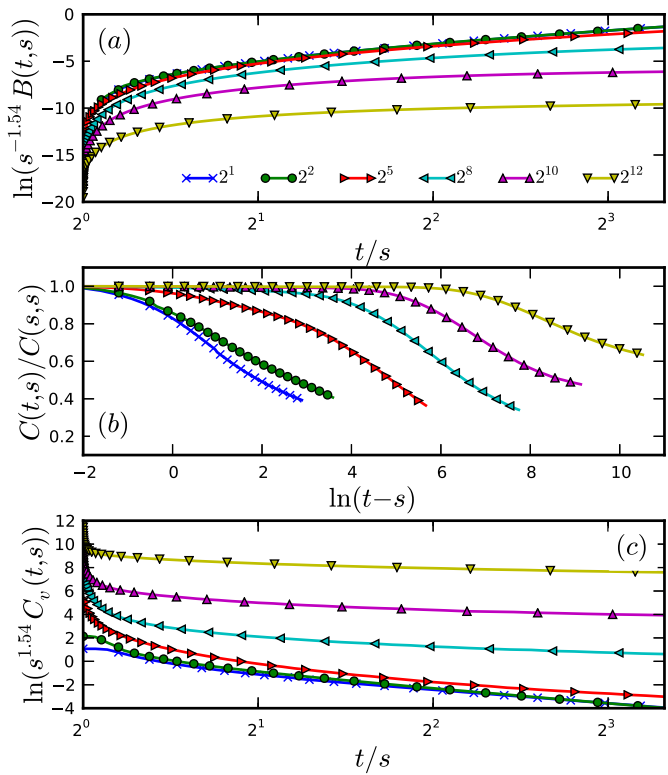}
  \caption{(Color online) Relaxation of (a) the mean-square displacement, 
    (b) the normalized height autocorrelation function, and (c) the density 
    autocorrelation in a system of interacting vortex lines with point pinning
    centers of strength $p=0.05\epsilon_0$  and $L=640$; data averaged 
    over $800$ realizations.}
  \label{fig:p_fullsystem}
\end{figure}
\begin{figure}[t]
  \centering
  \includegraphics[width=\columnwidth]{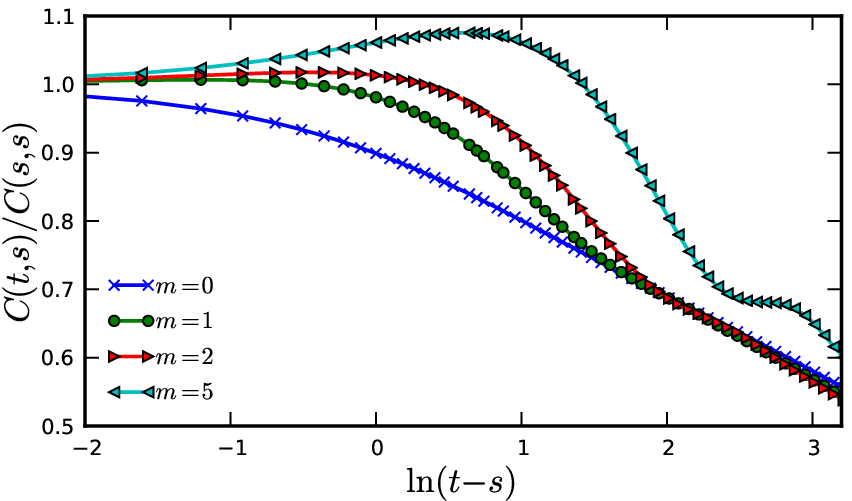}
  \caption{(Color online) The normalized two-time height autocorrelation 
    function $C(t,s)$ for different values of the vortex mass $m$ in a 
    system of interacting flux lines subject to randomly distributed point pins
    of strength $p=0.05\epsilon_0$, at waiting time $s=8$, and averaged 
    over $1000$ simulation runs. The data for $m>0$, for which $C(t,s)$ 
    displays non-monotonic and even oscillatory behavior, was generated 
    using the Br\"{u}nger-Brooks-Karplus 
    integrator~\cite{Izaguirre2001,Brunger1984}.}
  \label{fig:mass_variation}
\end{figure}
We are now in a position to investigate the system of interacting
vortex lines subject to attractive point-like disorder. As expected, the 
global time evolution, see Fig.~\ref{fig:relaxation-fullsystem}(a,b),
which is dictated by the mutual vortex repulsion, is hardly modified by the
defects. The effective exponent $\beta_B(t)$ displays essentially the 
same behavior as in the absence of pinning centers; compare
Figs.~\ref{fig:relaxation-fullsystem}(b) and 
\ref{fig:relaxation-interclean}(b). Similarly, the aging exponents and
the overall shapes of the mean-square displacement and the density-density
autocorrelation in Fig.~\ref{fig:p_fullsystem}(a,c) almost match the
simulation results without disorder. For short times $t<2^6$, the
effective gyration radius exponent $\beta_h(t)$ in 
Fig.~\ref{fig:relaxation-fullsystem}(d) is quite similar to $\beta_h(t)$ in the 
non-interacting case with point pins, see 
Fig.~\ref{fig:relaxation-noninter}(d). For longer times, repulsive forces
alter the relaxation of $r_g^2(t)$, which tends towards higher values, 
Fig.~\ref{fig:relaxation-fullsystem}(c). Hence, the global observables 
$B(t,s)$ and $C_v(t,s)$ are influenced mainly through the presence or 
absence of vortex-vortex repulsion through the ensuing mutual caging. 
The local quantity $C(t,s)$, on the other hand, better probes information on 
the disorder present in the sample.

In MC simulations~\cite{Pleimling2011}, an interesting, non-monotonic 
behavior was revealed in the height-height autocorrelation function for a
system of interacting vortex lines subject to point-like disorder: The height
autocorrelations displayed a pronounced maximum for small waiting times 
$s$ and $\ln(t-s)\approx5$. Yet this feature is absent when the 
corresponding system is investigated with our LMD algorithm; see 
Fig.~\ref{fig:p_fullsystem}(b). In the present study, we have taken the 
vortex mass per unit length to be small and neglected the inertial term in the 
Langevin equation; see Sec.~\ref{sec:lang-molec-dynam}. The algorithm 
used in Ref.~\cite{Pleimling2011} assumes a finite displacement per
MC step which generates an effective mass. This in turn gives rise to 
oscillatory behavior at short times. This interpretation is indeed confirmed by 
LMD simulations that allow for a mass term in the Langevin equation, as
depicted in Fig.~\ref{fig:mass_variation}. With increasing vortex mass, 
$C(t,s)$ shows damped oscillations on top of the monotonic time dependence 
of the overdamped, zero-mass case.

We observe two-step relaxation behavior, typical e.g. for structural glasses, 
in the normalized two-time height-height autocorrelation $C(t,s)$ when 
plotted as a function of the time difference $t-s$, shown in 
Fig.~\ref{fig:p_fullsystem}(b). This behavior was also reported in the 
previous MC study~\cite{Pleimling2011}: A $\beta$-relaxation regime 
where $C(t,s)$ is hardly changing and displays time translation invariance 
precedes the ultimate very slow decay. We attempted to fit a stretched 
exponential function to our long-time results, but could not achieve 
satisfactory agreement with our data.

\subsection{Non-interacting Vortices with Columnar Defects}
\label{sec:nonint-vort-lines}

\begin{figure}
  \centering
  \includegraphics[width=\columnwidth]{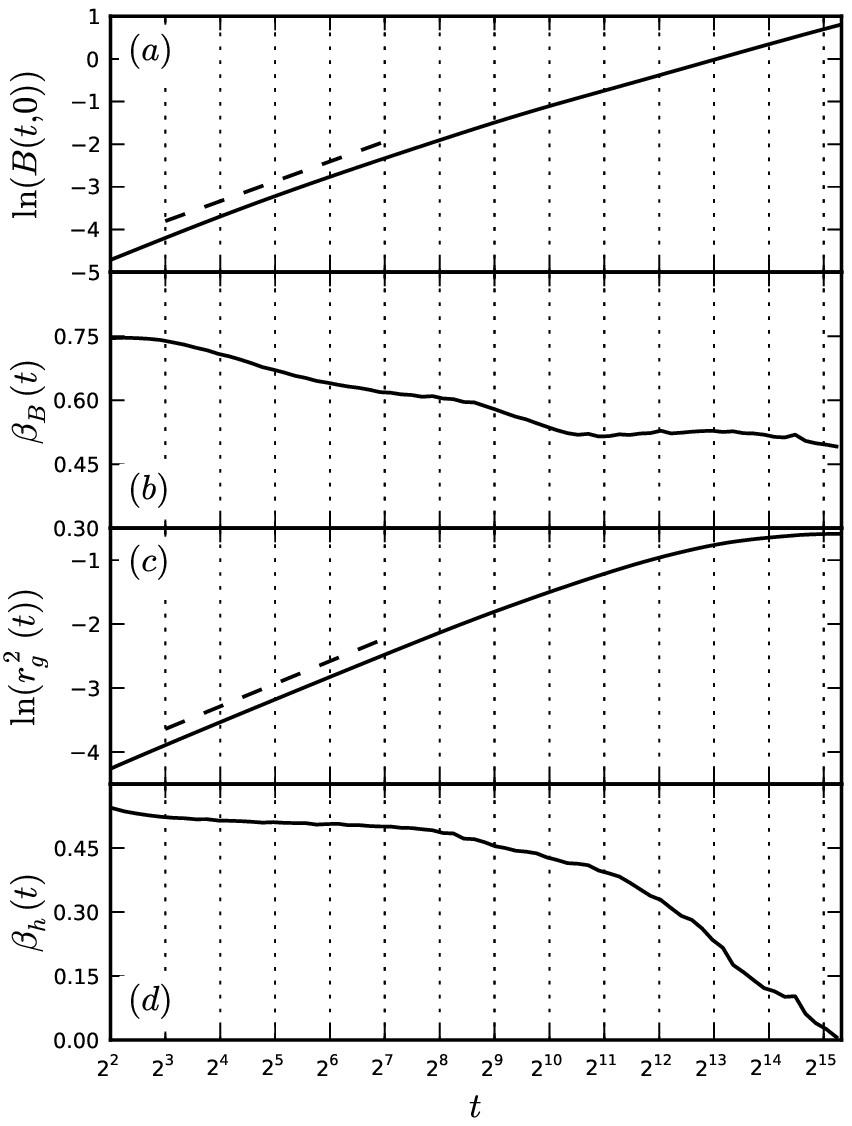}
  \caption{Relaxation behavior of (a) the flux line mean-square
    displacement $B(t,0)$, (c) the squared gyration radius $r_g^2(t)$,
    and (b, d) the associated effective exponents $\beta_B$ and
    $\beta_h$ over time for non-interacting vortices subject to
    columnar pinning centers with $p=0.05\epsilon_0$; data averaged
    over $10000$ realizations. The dashed lines indicate the power
    laws with the mean effective exponents
    $\overline{\beta_B}\approx0.672\pm0.037$ and
    $\overline{\beta_h}\approx0.510\pm0.006$ over the time range
    $2^3\le t\le2^7$.}
  \label{fig:c_relaxation_noninter}
\end{figure}
\begin{figure}
  \centering
  \includegraphics[width=\columnwidth]{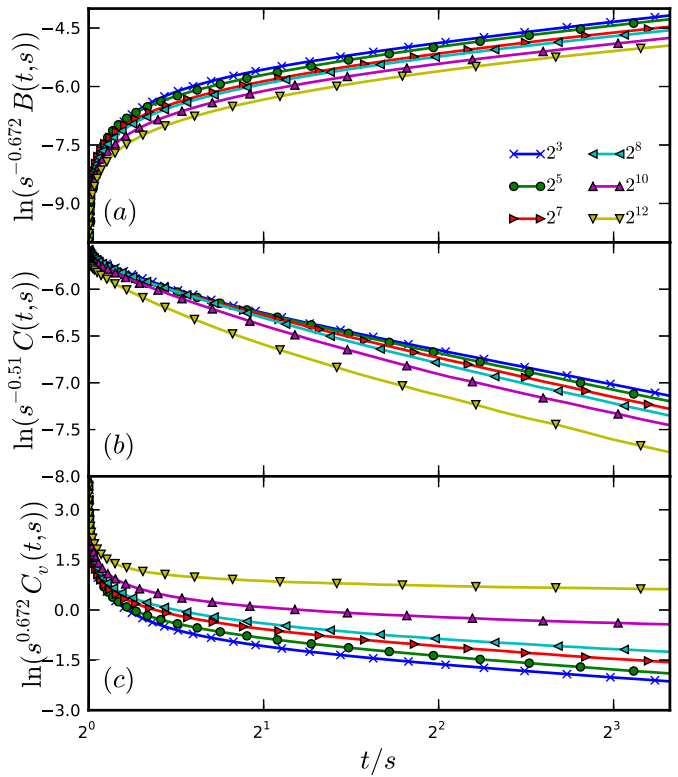}
  \caption{(Color online) Relaxation of (a) the mean-square
    displacement, (b) the height autocorrelation function, and (c) the 
    density autocorrelation in a system of non-interacting flux lines of
    length $L=640$ subject to randomly distributed columnar pins of 
    strength $p=0.05\epsilon_0$; data averaged over $1000$ realizations. 
    Scaling attempts with the averaged effective exponents taken from 
    Fig.~\ref{fig:c_relaxation_noninter} do not yield data collapse for 
    $B(t,s)$ and $C_v(t,s)$. One achieves better data collapse at early 
    waiting times $s$ for the height autocorrelations $C(t,s)$.}
  \label{fig:c_noninter}
\end{figure}
\tolerance=300
To compare the effects of uncorrelated point pins with those of
extended, correlated defects on the flux line relaxation kinetics, we
now investigate a system of non-interacting vortex lines in the
presence of columnar pinning centers with a pinning potential strength
$p=0.05\epsilon_0$. We start by first considering the case of
non-interacting flux lines relaxing in the presence of columnar
defects.  Figure~\ref{fig:c_relaxation_noninter} shows the relaxation
curves for $B(t,0)$ and $r_g^2(t)$ with their associated effective
exponents $\beta_B(t)$ and $\beta_h(t)$. The time evolution of the
mean-square displacement $B(t,0)$ is slightly accelerated compared to
the disorder\hyp{}free case for times up to $t=2^9$; see
Fig.~\ref{fig:relaxation-free-line}. This indicates that the initial
trapping of vortices at linear pinning sites happens during this time
regime. At later times, the effective exponent approaches the value of
free flux lines $\beta_B=0.5$, due to unbound vortex line
wandering. In fact, only $\sim 4$ to $5\%$ of the vortices are pinned
at the end of our simulation time window.

\tolerance=225 As mentioned above, the average number of pinning sites
per layer in our simulations is the same for point-like and columnar
defects. Hence the combined lateral cross section along the $z$ axis
of all pinning sites is much larger for point than it is for columnar
pins. Consequently, the probability for a vortex line to encounter a
randomly placed columnar pin during what is essentially a random walk
is much lower than in samples with randomly placed point defects. A
flux line is therefore most likely to be initially captured by a
single columnar defect (rather than multiple pinning sites, which
would lead to vortex kink configurations) along a short length span,
and subsequently becomes completely trapped at that pinning center;
this yields rather small values for the final gyration radius, see
Figs.~\ref{fig:c_relaxation_noninter}(c) and
\ref{fig:steady-state-ivgyrrad-columnar}(b). This is in stark contrast
with samples containing uncorrelated point defects, where a single
vortex line becomes captured by many pinning sites, which leads to a
larger terminal radius of gyration because the line is stretched in
random directions between multiple pins; see
Sec.~\ref{sec:nonint-lines-disorder}. Thus, point-like pinning centers
typically generate rough flux line configurations, whereas columnar
defects straighten bound vortices. This fact can be expressed as an
effective upward renormalization of the elastic line stiffness,
reflected macroscopically as a diverging tilt modulus for the entire
vortex system in the pinned Bose glass
phase~\cite{Nelson1993,Tauber1997}.

Comparing with the disorder-free system, the time evolution of the
effective exponent of the radius of gyration $\beta_h(t)$ in the
columnar defect case is also different: The exponent initially assumes
a value $0.5$ consistent with EW scaling. It begins to deviate from
the EW value at $t\approx2^9$, while in the non-disordered case this
decrease does not occur until $t\approx2^{12}$, see
Fig.~\ref{fig:c_relaxation_noninter}(d) and
Sec.~\ref{sec:free-nonint-lines}. This is due to the fact that thermal
line fluctuations inside a single columnar defect are confined to the
pinning potential well, which causes early saturation of the
equivalent EW growth process. An appreciable number of kinks and
double-kinks due to vortices trapped at multiple columnar pinning
sites would presumably alter this relaxation behavior, but the
occurrence probability of kinks is rather small in our system, as
explained above. At low temperatures, our non-interacting flux line
system in the presence of correlated disorder forms a very dilute Bose
glass, where the number of vortex lines is much less than the number
of pinning sites.

Figure~\ref{fig:c_noninter} shows the relaxation of the three two-time
autocorrelation functions for different waiting times $s$ as a function of 
the ratio $t/s$. We obtain data collapse for the height autocorrelation 
function $C(t,s)$ when scaling with the appropriately averaged effective 
aging exponent for early waiting times $s<2^9$ and $t/s<2$, which is 
consistent with the EW regime. The more global mean-square 
displacement $B(t,s)$ and the density autocorrelation $C_v(t,s)$ cannot 
similarly be scaled to obtain data collapse, because the effective 
exponent $\beta_B(t)$ is never even approximately constant througout 
the entire observed time interval, as is evident in 
Fig.~\ref{fig:c_relaxation_noninter}(b).

\subsection{Interacting Vortex Lines with Columnar Defects}
\label{sec:inter-vort-lines}

\begin{figure}
  \centering
  \includegraphics[width=\columnwidth]{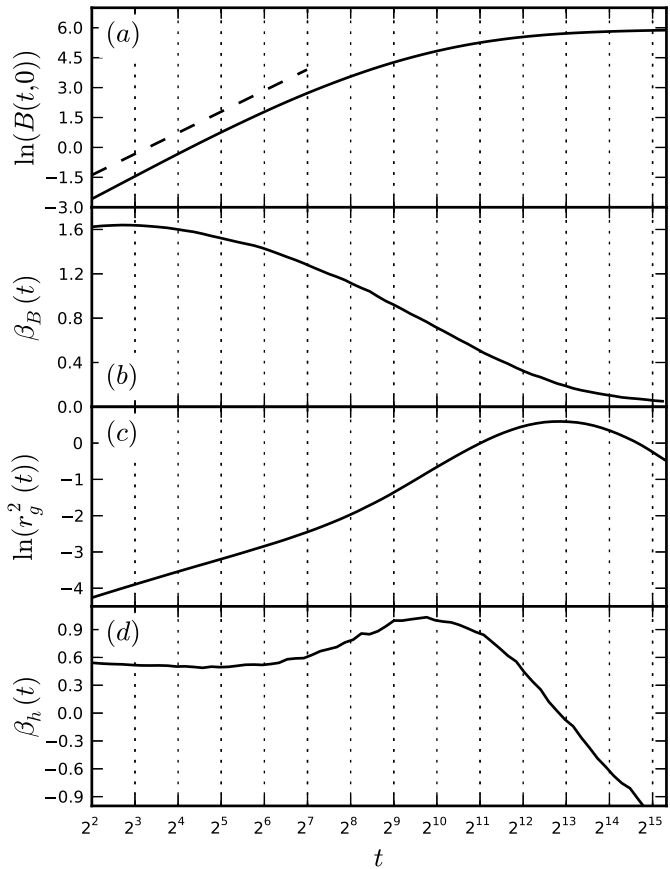}
  \caption{Non-equilibrium relaxation of (a) the flux line mean-square 
    displacement $B(t,0)$, (c) the squared gyration radius $r_g^2(t)$,
    and (b, d) the associated effective exponents $\beta_B$ and
    $\beta_h$ over time for interacting vortices in a system with 
    columnar defects of strength $p=0.05\epsilon_0$, averaged over 
    $1000$ realizations. The dashed line in (a) shows a power law with the 
    mean effective exponent $\overline{\beta_B}\approx1.53\pm0.11$, 
    averaged over the time interval $2^2\le t\le 2^7$.}
  \label{fig:c_relaxation_fullsystem}
\end{figure}
\begin{figure}
  \centering
  \includegraphics[width=\columnwidth]{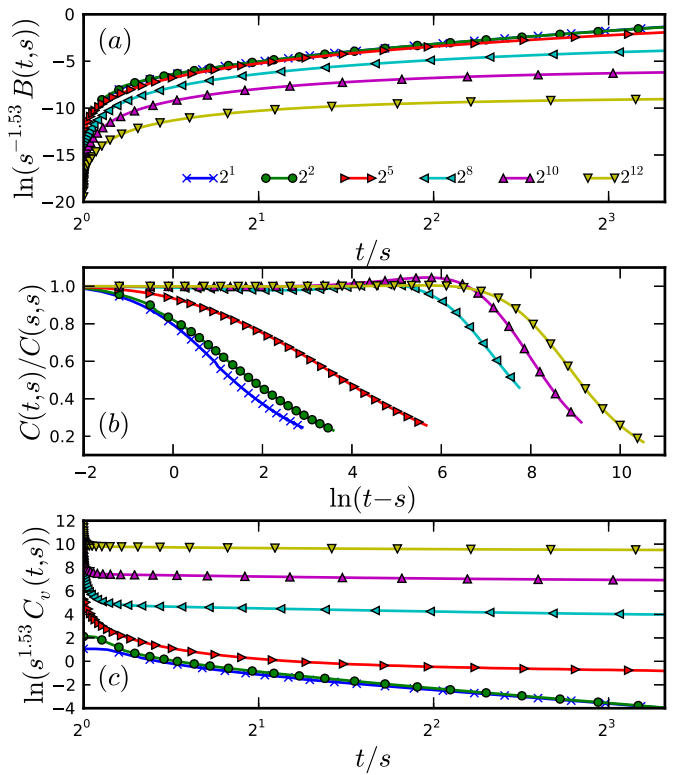}
  \caption{(Color online) Relaxation of (a) the mean-square
    displacement, (b) the normalized height autocorrelation function, and 
    (c) the density autocorrelation in a system of interacting vortices of
    length $L=640$ with columnar defects of strength $p=0.05\epsilon_0$; 
    data averaged over $800$ realizations.}
  \label{fig:c_fullsystem}
\end{figure}
Next, we again turn on the repulsive vortex-vortex interactions. Similar 
to the disorder-free system and the samples with point pinning centers, 
see Secs.~\ref{sec:free-inter-lines} and~\ref{sec:inter-lines-disorder}, 
caging effects accelerate vortex motion, and the shape of the single-time 
mean-square displacement $B(t,0)$ and its associated effective exponent 
are hardly modified by the disorder; compare 
Fig.~\ref{fig:c_relaxation_fullsystem}(a,b). At very large times 
$t>2^{12}$, $B(t,0)$ becomes flatter and approaches a plateau owing to 
the confinement of vortex lines by attractive defects. This effect is rather
more pronounced for columnar pins than for uncorrelated point defects. 
In fact, the maximal values of $B(t,0)$ and $r_g^2(t)$ are both smaller in 
the case of correlated disorder, which indicates tighter binding to the
pinning sites.

The radius of gyration also shows similar time evolution trends as 
compared to samples with point-like disorder; see 
Figs.~\ref{fig:c_relaxation_fullsystem}(c)
and~\ref{fig:relaxation-fullsystem}(c). The effects of attractive columnar
pinning sites set in later than for point-like pins, owing to the 
aforementioned differences in the encounter probability for flux lines
and pinning centers. The accelerated growth of the gyration radius for 
$t>2^9$ is due to the pinning at multiple sites and the subsequent 
formation of kinks, here facilitated by the strong repulsive forces. The 
non-monotonic behavior of $r_g^2(t)$ at times $t>2^{13}$ is caused by 
the decay of previously formed kinks. The density autocorrelation 
$C_v(t,s)$ in Fig.~\ref{fig:c_fullsystem}(c) becomes flat for waiting times 
$s>2^8$, which supports the interpretation that the vortex lines are
essentially trapped by this time and the only remaining relaxation process 
is the decay of metastable kink configurations.

Figure~\ref{fig:c_fullsystem}(b) depicts the normalized height-height
autocorrelation function of this system for different waiting times $s$. This 
function shows non-monotonic behavior, but in this situation it cannot stem 
from an effective mass, as we checked. The appearance of the maximum 
indicates a fundamental change in the lateral fluctuations. Although we do 
not yet fully understand this phenomenon, we tentatively relate this 
observation once again to the decay of kinks in the long-time limit.

\subsection{Finite-Size Effects}
\label{sec:finite-size-analysis}

\begin{figure}
  \centering
  \includegraphics[width=\columnwidth]{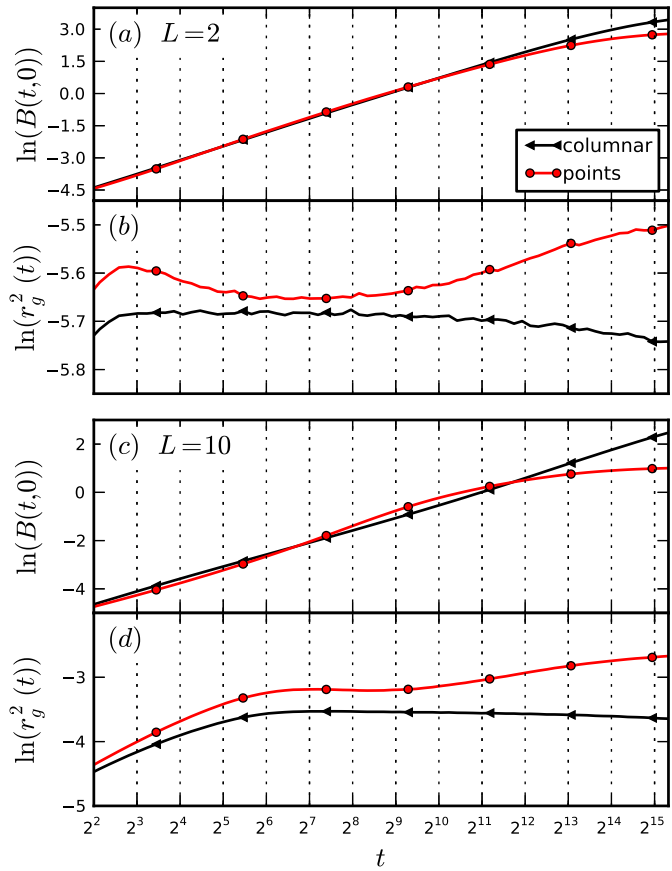}
  \caption{(Color online) Non-equilibrium relaxation of the flux line
    mean-square displacement $B(t,0)$ and the squared gyration radius
    $r_g^2(t)$ over time for non-interacting vortices in systems with
    point-like and columnar defects with identical pinning strengths
    $p=0.05\epsilon_0$ for (a,b) $L=2$ and (c,d) $L=10$, averaged over
    $10000$ realizations.}
  \label{fig:relaxation_thin_noninter}
\end{figure}
Effects due to the finite flux line length are best analyzed in terms of 
the second crossover between the EW and saturation regimes for 
non-interacting vortices in the absence of pinning sites; see
Sec.~\ref{sec:free-nonint-lines}.  (The first crossover between the
random thermal noise and the EW regimes only depends on the EW 
diffusion constant, which here corresponds to the vortex line tension.) 
The time at which this second crossover occurs depends on the square of 
the line length, $t_C=(L/24b_0)^22\pi/\epsilon_0$ (in the limit of large 
$L$)~\cite{Chou2009}. Our choice of $L=640$ for most of the simulation 
scenarios in this paper thus provides a sufficiently long time window 
$t_C\approx2^{12}$ to observe the competing effects of vortex 
interactions and pinning.

Of particular interest is the value of $L$ at which the relaxational
difference between point-like and columnar pinning sites becomes
apparent. Figure~\ref{fig:relaxation_thin_noninter} shows a comparison
plot of $B(t,0)$ and $r_g^2(t)$ for both columnar and point-like disorder 
for very short vortex lines with $L=2$ and $L=10$. In a purely 
two-dimensional system with $L=1$, where the flux lines are reduced to 
point particles, there is obviously no difference between the two types 
of pinning sites. But already for $L=2$ we observe differences in the 
long-time evolution of $B(t,0)$. The curve for columnar defects can be 
almost exactly reproduced by using point pins and halving the number of 
pinning sites. Hence the difference at $L=2$ is largely due to the lower 
effective density of columnar disorder, as discussed in 
Sec.~\ref{sec:nonint-vort-lines}. Yet this equivalence does not extend 
to the time evolution of the gyration radius $r_g^2(t)$, which displays 
small but significant qualitative differences for the two defect types 
throughout the simulation time window.

For $L=10$, the long-time difference in $B(t,0)$ between columnar and
point-like pinning sites can also be explained by the lower effective
density of columnar pinning sites. But the deviations appearing
at much shorter times reflect genuine physical distinctions in the pinning
behavior and ensuing relaxation kinetics. 

\section{Conclusion}
\label{sec:discussion}

In this paper, we have investigated the differences in the non-equilibrium
relaxation features between systems of magnetic flux lines in the 
presence of point-like and columnar disorder. Proceeding in a systematic 
way, and considering different limiting cases, allowed us to disentangle the 
distinct contributions originating from the attractive pinning centers, the 
repulsive mutual vortex interactions, and the line tension.

We validated both our Langevin Molecular Dynamics simulation code and 
the Monte Carlo algorithm used in previous studies in a genuine
out-of-equilibrium setting by comparing the steady-state vortex
velocity and radius of gyration as a function of an external driving
force to results from Monte Carlo simulations. As discussed in
Sec.~\ref{sec:comp-algor}, both these simulation methods need to be
tested and validated when applied to non-equilibrium situations. We
found that the pinning potential strength in MC is slightly
renormalized as compared to LMD due to the (inevitable) choice of a
maximal MC step size.

The introduction of columnar instead of point-like pinning sites
dramatically changes the steady-state properties. As expected, the 
critical depinning force is enhanced by approximately an order of
magnitude. The radius of gyration is suppressed via vortex line
confinement in columnar pinning sites for a driving force well below
the critical depinning force. At the transition, partial depinning
leads to the formation of half-loops and kinks in the vortex lines and
thus to a sharp increase in the radius of gyration. At even higher
driving forces, flux line motion is not influenced by pinning.

We carefully studied the relaxation towards equilibrium of a system of 
initially perfectly straight and randomly\hyp{}placed vortex lines under various 
conditions by observing single- and two-time quantities to again compare 
the effects of uncorrelated point pins and correlated extended defects,
and to further validate our LMD code against previously published MC
results. We investigated the possibility of data collapse and, hence, a 
simple aging scenario, by appropriately scaling our two-time quantities. 
We started with free, non-interacting vortex lines and showed that our
results completely agreed with the MC data and the predictions from
the Edwards-Wilkinson interface growth model. We then systematically
introduced attractive pinning centers and mutual repulsive vortex 
interactions. Caging effects due to vortex-vortex interactions lead to a 
considerable acceleration in the relaxation of global quantities, such as 
the single-time mean-square vortex displacement. A recent study 
revealed that the MC two-time height-height autocorrelation function 
for a system with interactions and point-like disorder displayed 
non-monotonic behavior (shown in Ref.~\cite{Pleimling2011}). 
Comparing with data obtained with an additional inertial term in our LMD 
algorithm, we argued that these oscillations stem from an effective mass 
generated by the introduction of a maximal MC step length.

We demonstrated that the relaxation behavior of vortex lines depends
crucially on the type of disorder. The vortex and Bose glass phases 
display complex non\hyp{}universal relaxation features that are highly 
dependent on the material parameters. Once a deeper understanding 
of the transient behavior has been established, detailed information 
contained in such time-dependent quantities could be used to 
characterize material properties and specific samples. Point-like disorder 
binds vortices to many pinning sites at once, while columnar defects 
capture entire flux lines. When comparing these two defect types, one 
needs to take into account the difference in the effective pin density 
and thus the distinct probability of vortex line elements to become 
trapped. One may characterize the vortex glass phase through the 
roughness exponent $\chi$ of the spatial height-height correlation 
function along the strongly fluctuating flux line trajectory. In contrast, 
correlated linear defects effectively enhance the elastic line stiffness 
and hence straighten the trapped vortices in the Bose glass phase.

We plan to expand our study to the transient properties of driven
vortex lines. The resulting relaxation then is towards a genuine
non-equilibrium state in contrast to the relaxation towards
equilibrium, which we presented in this paper. Other avenues for
further investigations are the study of different and more realistic 
initial conditions, such as magnetic field or temperature quenches.

\section*{Acknowledgements}
\label{sec:acknowledgements}
This research is supported by the U.S. Department of Energy, Office of 
Basic Energy Sciences, Division of Materials Sciences and Engineering
under Award DE-FG02-09ER46613.

\end{document}